\documentclass[twocolumn,english,reprint, longbibliography, superscriptaddress, breaklinks=true, showkeys, showpacs=false, nofootinbib]{revtex4}
\usepackage[T1]{fontenc}
\usepackage[latin9]{inputenc}
\usepackage{color}
\usepackage{babel}
\usepackage{amsmath}
\usepackage{amssymb}
\usepackage{graphicx}
\usepackage{physics}
\usepackage{subfigure}
\usepackage[unicode=true,pdfusetitle,
 bookmarks=true,bookmarksnumbered=false,bookmarksopen=false,
 breaklinks=true,pdfborder={0 0 0},backref=false,colorlinks=true]
 {hyperref} 
\usepackage{breakurl}
\usepackage{qcircuit}

\makeatletter
\@ifundefined{textcolor}{}
{%
 \definecolor{BLACK}{gray}{0}
 \definecolor{WHITE}{gray}{1}
 \definecolor{RED}{rgb}{1,0,0}
 \definecolor{GREEN}{rgb}{0,1,0}
 \definecolor{BLUE}{rgb}{0,0,1}
 \definecolor{CYAN}{cmyk}{1,0,0,0}
 \definecolor{MAGENTA}{cmyk}{0,1,0,0}
 \definecolor{YELLOW}{cmyk}{0,0,1,0}
}

\pdfoutput=1
\hypersetup{colorlinks=true,citecolor=blue,linkcolor=cyan,urlcolor=blue,filecolor= green, breaklinks=true}
\usepackage{url}
\usepackage{breakurl}

\makeatother

\begin{document}

\title{Predictability as a  quantum resource}

\author{Marcos L. W. Basso}
\email{marcoslwbasso@hotmail.com}
\thanks{corresponding author}
\address{Departamento de F\'isica, Centro de Ci\^encias Naturais e Exatas, Universidade Federal de Santa Maria, Avenida Roraima 1000, Santa Maria, Rio Grande do Sul, 97105-900, Brazil}
\address{New adress: Centro de Ci\^encias Naturais e Humanas, Universidade Federal do ABC, Avenida dos Estados 5001, 09210-580 Santo Andr\'e, S\~ao Paulo, Brazil}

\author{Jonas Maziero}
\email{jonas.maziero@ufsm.br}
\address{Departamento de F\'isica, Centro de Ci\^encias Naturais e Exatas, Universidade Federal de Santa Maria, Avenida Roraima 1000, Santa Maria, Rio Grande do Sul, 97105-900, Brazil}

\selectlanguage{english}%

\begin{abstract}
\textbf{Abstract}:
Just recently, complementarity relations (CRs) have been derived from the basic rules of Quantum Mechanics. The complete CRs are equalities involving quantum coherence, $C$, quantum entanglement, and predictability, $P$. While the first two are already quantified in the resource theory framework, such a characterization lacks for the last. In this article, we start showing that, for a system prepared in a state $\rho$, $P$ of $\rho$, with reference to an observable $X$, is equal to $C$, with reference to observables mutually unbiased (MU) to $X$, of the state $\Phi_{X}(\rho)$, which is obtained from a non-revealing von Neumann measurement (NRvNM) of $X$. We also show that $P^X(\rho)>C^{Y}(\Phi_{X}(\rho))$ for observables $X, Y$ not MU. Afterwards, we provide quantum circuits for implementing NRvNMs and use these circuits to experimentally test these (in)equalities using the IBM's  quantum computers. 
Furthermore, we give a resource theory for predictability, identifying its free quantum states and free quantum operations and discussing some predictability monotones.
Besides, after applying one of these predictability monotones to study bipartite systems, we discuss the relation among the resource theories of quantum coherence, predictability, and purity.
\end{abstract}


\maketitle

\section{Introduction}
Both Bohr's complementarity principle \cite{Bohr} and Heisenberg's uncertainty principle \cite{Robertson} were guidelines for the development of Quantum Mechanics. While the latter was formulated quantitatively since the genesis of Quantum Mechanics, we can safely say that the former was quantitatively formulated just recently, where the wave-particle duality appears as the main example of the Bohr's principle. In a two-way interferometer, such as the Mach-Zehnder interferometer or the double-slit interferometer, the wave aspect is characterized by the visibility of the interference fringes while the particle nature of the quanton \cite{Leblond} is given by the which-way information along the interferometer. The first quantitative version of the wave-particle duality was explored by Wootters and Zurek \cite{Wootters} from an information-theoretical approach, when they investigated two-slit interference in which one obtains incomplete which-way information by introducing a path-detecting device, and showed that a partial interference pattern can still be retained. Later, Greenberger and Yasin \cite{Yasin} formulated a quantitative relation given by an inequality expressed in terms of a priori information about the path, named predictability, and the interferometric visibility.

With the rapid development of the field of Quantum Information along with the increasing interest in Quantum Foundations, many approaches were taken for quantifying the wave-particle properties of a quantum system as well as to generalize such aspects for $d$-dimensional quantum systems with an arbitrary number of quantons \cite{Engle, Jaeger, Ribeiro, Bera, Coles, Hillery, Qureshi, Maziero, Lu}.  For instance, it was realized that the quantum coherence \cite{Baumgratz} can be considered as a natural generalization for the visibility of an interference pattern \cite{Bera, Bagan, Mishra}. With the axiomatization of Quantum Mechanics,
uncertainty relations that put Heisenberg's uncertainty principle on quantitative grounds were derived for any pair of observables, without appealing to a specific physical scenario. Following this line of thought, in Ref. \cite{Maziero} the authors explored the properties of the density matrix of a system $A$ to derive several complementarity relations of the type
\begin{align}
    C_{re}(\rho_A) + P_{vn}(\rho_A) \le \log_2 d_A, \label{eq:cr1}
\end{align}
where $C_{re}(\rho_A) := S_{vn}(\rho_{A diag}) - S_{vn}(\rho_{A})$ is the relative entropy of quantum coherence, $P_{vn}(\rho_A) = S_{vn}^{\max} - S_{vn}(\rho_{A diag})$ is the corresponding bone-fide predictability measure and $S_{vn}(\rho_A) = - \Tr \rho_A \log_2 \rho_A$ is the von Neumann entropy, with all complementarity measures regarded in Ref.  \cite{Maziero} satisfying the criteria established in Refs. \cite{Durr, Englert} for measures of visibility and predictability. As well, it was first noticed by Jakob and Bergou \cite{Janos} that, if one considers that the system $A$ is part of a bipartite (or multipartite) pure quantum system $AB$ such that $\rho_A = \Tr_B (\ket{\psi}_{AB}\bra{\psi})$, then the complementarity relation of the type of Eq. (\ref{eq:cr1}) can be completed \cite{Marcos, Leopoldo}:
\begin{align}
    C_{re}(\rho_A) + P_{vn}(\rho_A) + S_{vn}(\rho_A) = \log_2 d_A, \label{eq:ccr1}
\end{align}
where $S_{vn}(\rho_A)$ is an entanglement monotone which accompanies the relation above. Moreover, in Ref. \cite{Basso_ECCR}, the authors showed that for each complementarity measures that satisfies the criteria established in Refs. \cite{Durr, Englert}, it is possible to obtain an entanglement monotone that completes relations of the type of that in Eq. (\ref{eq:cr1}). Triality relations like Eq. (\ref{eq:ccr1}) are also named complete complementarity relations (CCRs), since the inequality is turned into an equality by taking into account the correlations with others systems, as was interpreted by Qian et al. in Ref. \cite{Qian}. Besides, the predictability measure together with the entanglement monotone can be considered as a measure of path distinguishability in an interferometer, as was recently discussed  in Ref. \cite{Tabish, Wayhs}, and first introduced by Englert in Ref. \cite{Engle}.

Another interesting feature of Eq. (\ref{eq:ccr1}) is that both quantum coherence and quantum entanglement are properties of quantum systems that can be seen as resources for certain tasks in the field of Quantum Information and Quantum Computation. This observation led to the formal development of quantum resource theories. A quantum resource theory is a formal framework that stems from the fact that certain properties of quantum systems become valuable resources for certain tasks under a restrictive set of operations \cite{Chit}. Perhaps, the major example of such quantum resources is entanglement, which is used for many quantum information processing tasks, when restricted to local operations and classical communication (LOCC) \cite{Hor}. Until now, the formal framework provided by resource theory has been applied to several properties of quantum systems such as purity \cite{Horo, Micha}, quantum coherence \cite{Baumgratz, Adesso}, quantum reference frames \cite{Spekk, Gour}, athermality in quantum thermodynamics \cite{Brandao}, contextuality \cite{Amaral}, quantum incompatibility \cite{Martins} and quantum (ir)reality \cite{Costa}. Besides, several approaches for a unifying framework of resource theories have been pursued lately \cite{Costa, Fernando, Lloyd}. Therefore, the natural question that arises when one looks carefully at Eq. (\ref{eq:ccr1}) is: Can predictability be characterized as a quantum resource? This is main the question that we propose to answer in this article.

Predictability is directly related to the capability of predicting what outcome will be obtained in a given run of an experiment. 
We will show, in Sec. \ref{sec:pred}, that the predictability with reference to an observable $X$ is equal to the quantum coherence with reference to observables mutually unbiased (MU) to $X$ of the state $\Phi_{X}(\rho)$, which is obtained from a non-revealing von Neumann measurement (NRvNM) of $X$. Therefore, with this we identify predictability with a well known quantum resource, partially answering the question above. 
After extending this result to observables not necessarily MU, in Sec. \ref{sec:exp} we provide quantum circuits to implement NRvNM and use these circuits to experimentally test these (in)equalities using the IBM quantum computers. Besides, in Sec. \ref{sec:resou} we report a resource theory for predictability (RTP), identifying its free states and operations and discussing some predictability monotones. After giving, in Sec. \ref{sec:geo}, a geometrical-intuitive description for the RTP, in Sec. \ref{sec:composite}, we apply one of the monotones to discuss predictability in the context of composite systems, and we address the relation among the resource theories of predictability, coherence, and purity in Sec. \ref{sec:RTs}.
Finally, in Sec. \ref{sec:con}, we give our concluding remarks.

\section{Predictability as a resource}
\label{sec:pred}
In this section, we will show that the predictability with reference to an observable $X$ is equal to the quantum coherence, with reference to observables MU to $X$, of the state $\Phi_{X}(\rho)$, which is obtained from a NRvNM of $X$ and we shall extend this result to observables that are not necessarily MU to $X$. To do this, we introduce the predictability from a operational point of view, which by its turn is motivated by the protocol used in Ref. \cite{Martins} to characterize the quantum incompatibility as a resource encoded in a physical context that allows the test  of the security for a given communication channel against information leakage.

Let us consider a quantum system $A$ which is described by a density operator $\rho_A \in \mathcal{D}(\mathcal{H_A)}$ of dimension $d_A$, where $\mathcal{D}(\mathcal{H_A)}$ is the set of all density operators over the Hilbert space $\mathcal{H}_A$, and an observable with discrete non-degenerate spectrum $X = \sum_j x_j X_j$, where $X_j = \ketbra{x_j}$. Therefore, if an unrevealed projective measurement is realized, we have the following quantum operation
\begin{align}
    \Phi_X(\rho_A) = \sum_j X_j \rho_A X_j = \sum_j \expval{\rho_A}{x_j}X_j,
\end{align}
which is a CPTP map \cite{Nielsen}. It is worth noticing that the state $\Phi_X(\rho_A)$ is diagonal in the eigen-basis of the observable $X$, i.e., $ \Phi_X(\rho_A) := \rho^X_{Adiag}$, what furnish us a probability distribution. Besides, if the spectrum of $X$ is degenerate, we have to consider $X$ as a complete set of compatible observables $X = \{X_1,..., X_n\}$, as defined in Ref. \cite{Laloe}. Let us remember that a complete set of compatible observables $X = \{X_1,..., X_n\}$ is defined by: (i) $[X_j, X_k] = 0\ \forall j \neq k$ for $j,k = 1, ..., n.$ (ii) for each $n$-tuple of eigenvalues $\{x^1_j, ..., x^n_j \}$  of the observables $X_1,..., X_n$, we have a unique eigenvector. In other words,  a complete set of compatible observables $X = \{X_1,..., X_n\}$ is defined by a unique orthonormal basis of common eigenvectors. Therefore, the procedure of non-revealed projective measurements is given by
\begin{align}
    \Phi_X(\rho_A) =  \Phi_{X_1,...,X_n}(\rho_A) = \sum_j X_j \rho_A X_j,
\end{align}
where the eigenvector $\ket{x_j}$ is related to a particular $n-$tuple of eigenvalues $\{x^1_j, ..., x^n_j \}$. Besides, since $X_1,..., X_n$ are compatible, $\Phi_{X_j, X_k}(\rho_A) = \Phi_{X_k, X_j}(\rho_A)$, i.e., $\Phi_{X_j} \Phi_{X_k}(\rho_A) = \Phi_{X_k} \Phi_{X_j}(\rho_A)\ \forall j,k = 1,..., n$ and the particular order in which we perform the non-revealed measurements of the compatible observables does not matter. Therefore our analysis and results can be readily extended for the case where the observables are degenerate.  Now, let us consider another discrete  non-degenerate spectrum observable $Y = \sum_k y_k Y_k = \sum_k y_k \ketbra{y_k}$, which is maximally incompatible with $X$, i.e., $[X,Y] \neq 0$ such that $\abs{\braket{x_j}{y_k}}^2 = 1/d_A \ \forall j,k$. By performing a non-selective measurement of $Y$ given the state $\Phi_X(\rho_A)$, a straightforward calculation shows that $\Phi_{YX}(\rho_A):=\Phi_Y\Phi_X(\rho_A) = I_A/d_A$, being basis-independent, where $I_A$ is the identity operator on $\mathcal{H}_A$. Such map will be introduced in the Sec. \ref{sec:resou} as a maximally resource-destroying map \cite{Lloyd}. Now, since the von Neumann entropy of $\Phi_{YX}(\rho_A)$ is maximal, i.e., $S_{vn}(\Phi_{YX}(\rho_A)) = S_{vn}^{\max}$, we can rewrite the predictability measure with respect to the observable $X$ as
\begin{align}
    P^X_{vn}(\rho_A)& = S_{vn}^{\max} - S_{vn}(\rho^X_{Adiag}) \\
    & = S_{vn}(\Phi_{YX}(\rho_A)) - S_{vn}(\Phi_X(\rho_A)), \label{eq:pvn}
\end{align}
which gives us an operational definition for $P^X_{vn}(\rho_A)$. Besides, the predictability measure can be recast as the distinguishability between $\Phi_X(\rho_A)$ and $\Phi_{YX}(\rho_A)$: $P^X_{vn}(\rho_A) := S_{vn}(\Phi_X(\rho_A)||\Phi_{YX}(\rho_A)) = S_{vn}(\rho^X_{Adiag}||I_A/d_A)$, where $S_{vn}(\rho||\sigma):= \Tr (\rho \log_2 \rho -  \rho \log_2 \sigma)$ is the relative-entropy \cite{Nielsen}. This gives us a very nice interpretation for the predictability measure, i.e., $P^X_{vn}(\rho_A)$ measures how much the probability distribution given by $\Phi_X(\rho_A)$ differs from the uniform probability distribution $I_A/d_A$. Finally, it's worth mentioning that $P^X_{vn}(\rho_A)$ is equal to the measure of quantum incompatibility for MU observables, which is a type of quantum resource \cite{Martins}.

By considering the relative entropy of quantum coherence (REQC) of a given state $\rho$ \cite{Baumgratz}, i.e., $C_{re}(\rho) = \min_{\iota \in I} S_{vn}(\rho||\iota)
= S_{vn}(\rho_{diag}) - S_{vn}(\rho)$, where $I$ is the set of all incoherent states. It is straightforward to see that the REQC with reference to the observable $Y$ of the state $\Phi_X(\rho_A)$ is given by
\begin{align}
    C^Y_{re}(\Phi_X(\rho_A)) = S_{vn}(\Phi_{YX}(\rho_A)) - S_{vn}(\Phi_{X}(\rho_A)),
\end{align}
which implies the following equality
\begin{equation}
    P_{vn}^{X}(\rho_A) = C_{re}^{Y}(\Phi_{X}(\rho_A)).
    \label{eq:PC}
\end{equation}
Therefore, the quantum coherence with reference to an observable $Y$ encoded in the state $\Phi_X(\rho_A)$, where $X,Y$ are MU, is equal to the predictability regarding the observable $X$ given the state $\rho_A$. Thus, if $P_{vn}^{X}(\rho_A)$ reaches its maximal possible value, so does $C_{re}^{Y}(\Phi_{X}(\rho_A))$. It is worth to stress here the operational point of view of this equality. Given a quanton $A$ described by a state $\rho_A$, by performing NRvNM with respect to observable $X$, we obtain a probability distribution $\Phi_X(\rho_A)$ which we can compare to the uniform probability distribution $I_A/d_A$ through the predictability measure $P^X_{vn}(\rho_A)$. In turns, $P^X_{vn}(\rho_A)$ dictates how much quantum coherence, with respect to the observable $Y$, we will have encoded in the state $\Phi_X(\rho_A)$.

On the other hand, if the observables $X,Y$ are \textit{not} MU, then it is easy to see that
\begin{align}
    C^Y_{re}(\Phi_X(\rho_A)) & = S_{vn}(\Phi_{YX}(\rho_A)) - S_{vn}(\Phi_{X}(\rho_A)) \\
   & = P^X_{vn}(\rho_A) - P^Y_{vn}(\Phi_X(\rho_A))\\
   & \le P^X_{vn}(\rho_A).
   \label{eq:CPineq}
\end{align}
That is to say, the amount of quantum coherence with reference to the observable $Y$ encoded in state $\Phi_X(\rho_A)$ will be smaller by the amount given by the predictability regarding $Y$ of the state $\Phi_X(\rho_A)$. The equation above can still be recast as
\begin{align}
    C^Y_{re}(\Phi_X(\rho_A)) +  P^Y_{vn}(\Phi_X(\rho_A)) =  P^X_{vn}(\rho_A),
\end{align}
which implies that the complementarity relation with respect to the observable $Y$ for the state $\Phi_X(\rho_A)$ is controlled by the predictability measure regarding the observable $X$ before we do any non-selective measurement on the state $\rho_A$. Besides, if $X,Y$ are MU, it is easy to see that we recover Eq. (\ref{eq:PC}). Now, by noticing that $P^X_{vn}(\rho_A) = P^X_{vn}(\Phi_X(\rho_A))$, one can easily see that the quantum incompatibility measure, introduced in Ref. \cite{Martins}, can be recast as the difference $\mathfrak{I}_{X,Y}(\rho_A) = P^X_{vn}(\Phi_X(\rho_A)) - P^Y_{vn}(\Phi_X(\rho_A))$, i.e., as the difference between the predictability related to the observable $X$ and the predictability related to the observable $Y$ encoded in the state $\Phi_X(\rho_A)$. For instance, if $[X,Y] = 0$, then $\mathfrak{I}_{X,Y}(\rho_A) = 0$, since the observables $X,Y$ share the same eigen-basis, and therefore have the same predictability given the state $\Phi_X(\rho_A)$. Finally, we have the following relation between complementarity relation of different observables: $ C^Y_{re}(\Phi_X(\rho_A)) +  P^Y_{vn}(\Phi_X(\rho_A)) =  P^X_{vn}(\Phi_X(\rho_A)) + C^X_{re}(\Phi_X(\rho_A)).$ Once the sum $ C_{re}(\rho_A) + P_{vn}(\rho_A)$ is invariant against a change of basis, we can drop $\Phi_X$ to obtain
\begin{align}
C^Y_{re}(\rho_A) + P^Y_{vn}(\rho_A) = C^X_{re}(\rho_A) + P^X_{vn}(\rho_A).   
\end{align}

Another interesting aspect of predictability measures is that it is possible to derive complementarity relations between the predictability of two observables $X,Y$ given the state $\rho_A$ from the entropic uncertainty relations. Let us consider \cite{Maassen}
\begin{align}
    H_X(\rho_A) + H_Y(\rho_A) \ge \log_2 \frac{1}{c}, \label{eq:unc}
\end{align}
where $H_X(\rho_A) = \sum_j p_j(X, \rho_A) \log_2 p_j(X, \rho_A)$, with $p_j(X, \rho_A) = \Tr (X_j \rho_A)$, is the Shannon's entropy of the observable $X$ given the state $\rho_A$, and similarly for $Y$. Meanwhile, $c = \max_{j,k} \abs{\braket{x_j}{y_k}}^2$, where $\ket{x_j}$ and $\ket{y_k}$ are the eigenvectors of the observables $X$ and $Y$, respectively. Therefore, by noticing that $P^X_{vn}(\rho_A) = \log_2 d_A - H_X(\rho_A)$ and similarly for $P_{vn}^Y(\rho_A)$, we see that Eq. (\ref{eq:unc}) can be rewritten as 
\begin{align}
    P^X_{vn}(\rho_A) + P^Y_{vn}(\rho_A) \le 2 \log_2 d_A + \log_2 c,
\end{align}
which is a complementarity relation between the predictability of the observables $X$ and $Y$ given the state $\rho_A$. Now, if the observables $X,Y$ are MU, we have $c = 1/d_A$, which implies that $P^X_{vn}(\rho_A) + P^Y_{vn}(\rho_A) \le \log_2 d_A.$

\section{Experimental verification of predictability-coherence (in)equalities}
\label{sec:exp}

In this section, we shall use IBM's quantum computers \cite{ibmq} to verify experimentally the predictability-coherence (in)equalities obtained in the previous section. We will start by giving a quantum circuit to perform non-revealing von Neumann quantum measurements on a qubit, and afterwards we generalize this circuit for an arbitrary number of qubits. These quantum circuits are then implemented on a real hardware for one- and two-qubit NRvNMs.

\subsection{One qubit}

\begin{figure}[t]
$\Qcircuit @C=1em @R=.7em { 
\lstick{\rho} & \gate{U^{\dagger}(\theta,\phi,\lambda)} & \ctrl{1} & \gate{U(\theta,\phi,\lambda)} & \qw & \rstick{\Pi_{n_{0},n_{1}}(\rho)} \\
\lstick{\ket{0}} & \qw & \targ & \qw & \qw
}$
\caption{Quantum circuit for implementing an one-qubit non-revealing von Neumann quantum measurement on a quantum computer. We use $V=U(\theta,\phi,\lambda)$ with $\lambda=0$ and $U^{\dagger}(\theta,\phi,\lambda)=U(\theta,\pi-\lambda,-\pi-\phi)$  with $U(\theta,\phi,\lambda)=\begin{bmatrix} \cos(\theta/2) & -e^{i\lambda}\sin(\theta/2) \\ e^{i\phi}\sin(\theta/2) & e^{i(\phi+\lambda)}\cos(\theta/2) \end{bmatrix}$.}
\label{fig:1qbqc}
\end{figure}
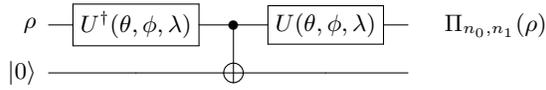

We want to perform a non-selective von Neumann measurement of a general one-qubit observable $\hat{n}\cdot\vec{\sigma}$ using unitary operations on this qubit and on an auxiliary system. Let us write the eigenbasis of this observable as 
\begin{align}
& |n_{0}\rangle = \cos(\theta/2)|0\rangle + e^{i\phi}\sin(\theta/2)|1\rangle = V|0\rangle, \\
& |n_{1}\rangle = -\sin(\theta/2)|0\rangle + e^{i\phi}\cos(\theta/2)|1\rangle = V|1\rangle,
\end{align}
with $V = \begin{bmatrix} \cos(\theta/2) & -\sin(\theta/2) \\ e^{i\phi}\sin(\theta/2) & e^{i\phi}\cos(\theta/2) \end{bmatrix}$.
First, we notice that the Controlled-NOT operation implements a non-revealing measurement on the computational basis, i.e., for $|\Psi\rangle_{AB} = CNOT_{A\rightarrow B}|\psi\rangle_{A}\otimes|0\rangle_{B}$ we have $Tr_{B}(|\Psi\rangle_{AB}\langle\Psi|) = \sum_{j=0}^{1}P_{j}|\psi\rangle_{A}\langle\psi|P_{j} \equiv \Pi_{0,1}(|\psi\rangle)$ with $P_{j}=|j\rangle\langle j|$ and $Tr_{B}$ is the partial trace operation \cite{ptrace}. Then, we see that 
$\Pi_{n_{0},n_{1}}(|\psi\rangle) = \sum_{j=0}^{1}P_{n_{j}}|\psi\rangle\langle\psi|P_{n_{j}} = V\Pi_{0,1}(|\phi\rangle)V^{\dagger}$ with $P_{n_{j}} = |n_{j}\rangle\langle n_{j}|$ and $|\phi\rangle = V^{\dagger}|\psi\rangle$. So, by the linearity of quantum dynamics, we see that non-selective measurements of a qubit observable $\hat{n}\cdot\vec{\sigma}$, of a system prepared in the state $\rho$, can be implemented using the quantum circuit shown in Fig. \ref{fig:1qbqc}.

\begin{table}
\caption{\label{tb:belem} Parameters for the ibmq\textunderscore belem chip.}
\begin{tabular}{l c c}
\hline 
Calibration parameters & Q0 & Q1 \tabularnewline
\hline 
\hline 
Frequency (GHz) & 5.09 & 5.245 \tabularnewline
T1 ($\mu$s) & 101.37 & 91.23 \tabularnewline
T2 ($\mu$s) & 118.44 & 79.31 \tabularnewline
Readout error ($10^{-2}$) & 2.14 & 1.83 \tabularnewline
CNOT error ($10^{-2}$) & $0\_1:1.903$ & $1\_0:1.903$ \tabularnewline
\hline
\end{tabular}
\end{table}

\begin{figure}[t]
\includegraphics[scale=0.55]{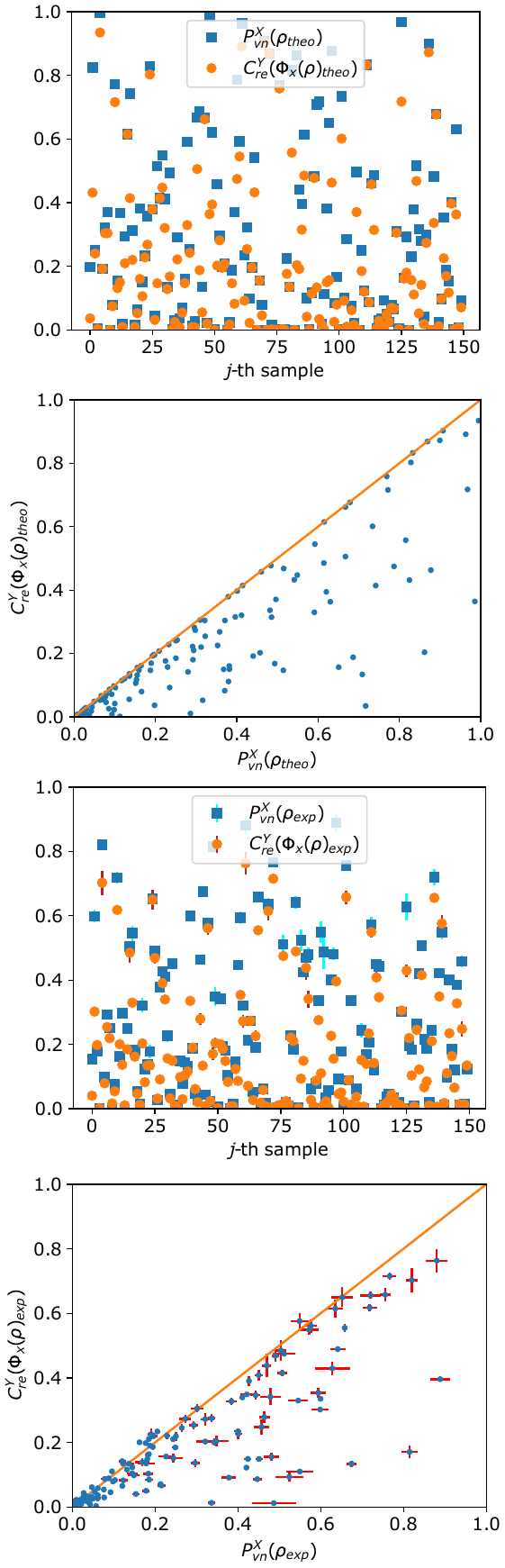}
\caption{(Color online) Theoretical and corresponding experimental values of the predictability of $\rho$, with reference to $X$, and of the quantum coherence of $\Phi_{X}(\rho)$, with respect to $Y$, for $150$ randomly chosen pure states $\rho=|\psi\rangle\langle\psi|$ and one-qubit observables $X$ and $Y$. The error bars are the standard deviation for four repetitions of the experiment.}
\label{fig:rand_1qb}
\end{figure}

The theoretical and the corresponding experimental results are shown in Fig. \ref{fig:rand_1qb}. For these experiments, we used the Belem chip, whose calibration parameters are shown in Table \ref{tb:belem}. We have randomly chosen 150 trios $(\rho=|\psi\rangle\langle\psi|,X,Y)$. The experimental results allow us to say that the inequality in Eq. (\ref{eq:CPineq}) is satisfied generally. The exceptions, for low values of $C$ and $P$, can be understood as being due to the creation of coherence from highly mixed states, that have origin on the systematic hardware errors, mainly gate errors, once this issue is not present in the simulated results (a similar issue appeared in Refs. \cite{Pozzobom, Mauro}).

\subsection{n qubits}
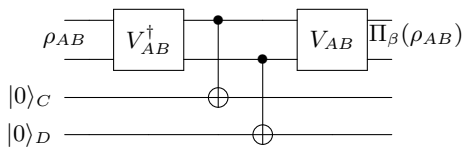
\begin{figure}[htp]
$
\Qcircuit @C=1em @R=.7em {
& \qw & \multigate{1}{V_{AB}^{\dagger}} & \ctrl{2} & \qw & \multigate{1}{V_{AB}} & \qw \\
\ustick{\rho_{AB}} & \qw & \ghost{{V_{AB}^{\dagger}}} & \qw & \ctrl{2} & \ghost{{V_{AB}}} & \qw  &  \ustick{\Pi_{\beta}(\rho_{AB})} \\
\lstick{|0\rangle_{C}}  & \qw & \qw & \targ & \qw & \qw  & \qw \\ 
\lstick{|0\rangle_{D}} & \qw & \qw & \qw & \targ & \qw  & \qw 
}
$
\caption{Quantum circuit to perform non-revealing von Neumann measurements $\Pi_{\beta}(\rho_{AB})$ in the general two-qubit basis $\beta = \left\{|\beta_{j,k}\rangle_{AB}=V|j\rangle_{A}\otimes|k\rangle_{B}\right\}_{j,k=0}^{1}$. }
\label{fig:2qbqc}
\end{figure}

Let us start with the two-qubit case. In view of the development made in the previous subsection, one can see that if we use two auxiliary qubits $C$ and $D$ and two controlled NOT operations $CNOT_{A\rightarrow C}$ and $CNOT_{B\rightarrow D}$, we can utilize $|\tau\rangle = CNOT_{B\rightarrow D}CNOT_{A\rightarrow C}|\Psi\rangle_{AB}\otimes|00\rangle_{CD}$ to implement a NRvNM in the computational basis: $\Pi_{\kappa}(|\Psi\rangle_{AB}) = Tr_{CD}(|\tau\rangle\langle\tau|)$, with $\kappa=\{|jk\rangle_{AB}\}_{j,k=0}^{1}$. Above and hereafter, we use the notation $|jk\rangle\equiv|j\rangle\otimes|k\rangle$. A NRvNM in a general two-qubit basis $\beta = \{|\beta_{j,k}\rangle = V|jk\rangle\}_{j,k=0}^{1}$, with $V$ being a unitary matrix, can be written as follows $\Pi_{\beta}(\rho)=\sum_{j,k}|\beta_{j,k}\rangle\langle\beta_{jk}|\rho|\beta_{j,k}\rangle\langle\beta_{jk}| = V\Pi_{\kappa}(\tilde{\rho})V^{\dagger}$, where $\tilde{\rho}=V^{\dagger}\rho V$. So, the quantum circuit to implement this general two-qubit NRvNM is shown in Fig. \ref{fig:2qbqc}.

\begin{figure}[ht]
    \centering
    \includegraphics[scale=0.6]{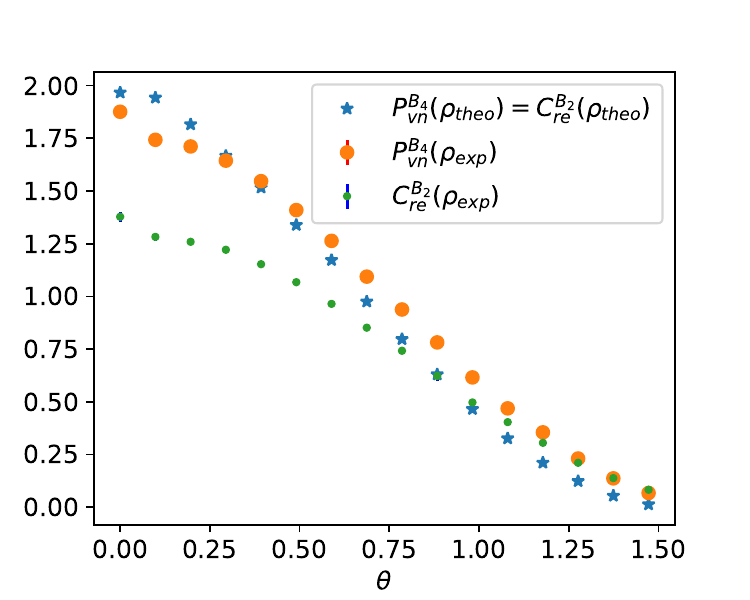}
    \caption{Theoretical and corresponding experimental values of the predictability of $|\Psi\rangle_{AB}$ of Eq. (\ref{eq:psi2qb}), with reference to the standard basis $\kappa\equiv B_{4}$ (so $V_{AB}=\mathbb{I}_{AB}$), and of the quantum coherence of $\Phi_{\kappa}(|\Psi\rangle_{AB}\langle\Psi|)$, with respect to mutually unbiased basis $B_{2}$ (Eq. (\ref{eq:B2})). The error bars are the standard deviation for four repetitions of the experiment.}
    \label{fig:PC2qb}
\end{figure}

Here we shall use this quantum circuit to verify the predictability-coherence equality for the two-qubit state 
\begin{align}
\ket{\Psi}_{AB} &=\cos^{2}(\theta/2) \ket{00} +\sin^{2}(\theta/2)\ket{11} \nonumber \\
& \hspace{0.3cm} +\sin(\theta/2)\cos(\theta/2)(\ket{01}+\ket{10}).
\label{eq:psi2qb}
\end{align}
For these experiments we utilized the Belem chip with averaged calibration parameters: $\text{CNOT Error}=
1.119\text{x}10^{-2}$, $\text{Readout Error}=
2.348\text{x}10^{-2}$, $T_{1}=
86.85\ \mu\text{s}$, and $T_{2}=
102.76\ \mu\text{s}$.
The obtained results are shown in Fig. \ref{fig:PC2qb}. The predictability is computed with reference to the computational basis $\kappa\equiv B_{4}$ and the quantum coherence is computed with reference to the basis (see Ref. \cite{mubs}):
\begin{equation}
B_{2} = \left\{
(|0_{1}0_{0}\rangle\pm i|1_{1}1_{0}\rangle)/\sqrt{2}, \\
(|1_{1}0_{0}\rangle\pm i|0_{1}1_{0}\rangle)/\sqrt{2}\right\},
\label{eq:B2}
\end{equation}
which is mutually unbiased to $\kappa$. Above $|0_{0}\rangle=(|0\rangle+|1\rangle)/\sqrt{2}$, $|1_{0}\rangle=(|0\rangle-|1\rangle)/\sqrt{2}$, $|0_{1}\rangle=(|0\rangle+i|1\rangle)/\sqrt{2}$, and $|1_{0}\rangle=(|0\rangle-i|1\rangle)/\sqrt{2}$.

In Fig. \ref{fig:PC2qb}, we see that there is a fairly good agreement between the theoretical prediction and the experimental results. The lower than expected values for the quantum coherence are due to systematic hardware errors, once the error bars are quite small and its associated quantum circuit has greater depth than that for the predictability, whose values are well closer to what is theoretically expected.

Finally, we must mention that the extension of quantum circuit of Fig. \ref{fig:2qbqc} for $n$ qubits is straightforward. Using $n$ auxiliary qubits and $n$ CNOT gates we can make a NRvNM in the computational basis. A NRvNM in an arbitrary $n$-qubit basis can then be obtained using the associated $n$-qubit unitary transformation $V$.

\section{Resource theory of predictability}
\label{sec:resou}
A resource theory relies on the definition of the following basic elements: (i) free states, (ii) free operations, (iii) resource states together with resource measures (monotones). These elements are closely related to each other, since, within this formal structure, free operations  should  not  be  able  to create the resource  from  free states, and resource measures should be monotonically non-increasing under free operations \cite{Chit}. Therefore, in this section we formalize the notion of a resource theory of predictability (RTP), identifying its free states and free operations and discussing some predictability monotones. To accomplish this task, we follow Ref. \cite{Lloyd}, where the authors developed a unified framework based on resource-destroying maps for defining resource theories. These maps have the property of leaving free states unchanged or taking free states to free states and erasing all the resource from resource states.

Let us consider again a quantum system $A$ described by a density operator $\rho_A \in \mathcal{D}(\mathcal{H_A)}$ of dimension $d_A$, where $\mathcal{D}(\mathcal{H_A)}$ is the set of all density operators over the Hilbert space $\mathcal{H}_A$, and an observable with discrete spectrum $X = \sum_j x_j X_j$, where $X_j = \ketbra{x_j}$. Therefore, our reference basis is given by the eigenvectors of the observable $X$. To define the set of free states for predictability with respect to the observable $X$, we remember that any bona fide predictability measure, as our prototypical measure defined in Eq. (\ref{eq:cr1}), should vanish when all the outcomes of the observable $X$ are equally probable \cite{Durr, Englert}, i.e., when $\expval{\rho_A}{x_j} = 1/d_A \ \forall j$. Therefore, it is easy to see that the set of free states with respect to the observable $X$, i.e., the set of maximally unpredictable states, denoted here by $\Upsilon_X(\mathcal{H_A})$, is the set of states of the type
\begin{align}
    \upsilon_A = & \frac{1}{d_A}\Big(\sum_j \ketbra{x_j} + \sum_{j \neq k} \epsilon_{jk} \ketbra{x_j}{x_k}\Big)\\
    & = (1 - p)\frac{I_{A}}{d_A}  + p \ketbra{\psi_d} \label{eq:mmc}
\end{align}
with $\epsilon_{jk} = p e^{i (\phi_j - \phi_k)} \in \mathbb{C}$ such that $\abs{\epsilon_{jk}} \le 1$ and $p \in [0,1]$; and $\ket{\psi_d} = \frac{1}{\sqrt{d}}\sum_j e^{i \phi_j} \ket{x_j}$. From Eq. (\ref{eq:mmc}), one can see that the set of unpredictable states consists of maximally coherent mixed states, which where introduced in Ref. \cite{Dhar}. Equivalently, $\Upsilon_X(\mathcal{H_A}):=\{\upsilon_A \in \mathcal{D}(\mathcal{H_A)} \ \text{such that}  \ \expval{\upsilon_A}{x_j} = 1/d_A \ \forall j \}$. It is noteworthy that when $\abs{\epsilon_{jk}} = 1$, $\upsilon_A$ is a maximally coherent pure state, which is a free state within the framework of a RTP with respect to the observable $X$. However, it is a resource state when we consider the resource theory of quantum coherence with reference to the observable $X$. Besides, let us recall the operational definition of pure states \cite{Nielsen}: for any pure state, should exist an observable $Y$ such that its outcome is completely predictable. Therefore, with the respect to the eigenbasis of the observable $Y$, such maximally coherent pure state is a resourceful state within the RTP. Furthermore, from the operational definition of our prototypical predictability measure defined in Sec. \ref{sec:pred}, a predictability measure does not distinguish, for instance, maximally coherent pure states from maximally incoherent states, since we have to perform NRvNM $\Phi_X$ to obtain the probability distribution $\Phi_X(\upsilon_A)$. This is in complete agreement with the criteria established in  Refs. \cite{Durr, Englert}, that asserts that any bone-fide predictability measure should be a \textit{continuous} function \textit{only} of the diagonal elements of $\rho_A$. The other predictability monotones that we will discuss later also satisfy the criteria established in the literature. Below, we prove that the set $\Upsilon_X(\mathcal{H_A})$ is convex and compact with respect to the observable $X$.

First, we proof that the set $\Upsilon_X(\mathcal{H_A})$ is \textit{convex}. Let $\upsilon_1, \upsilon_2 \in \Upsilon_X(\mathcal{H_A})$. Expressing $\upsilon_1, \upsilon_2$ in the eigenbasis of $X$: $\upsilon_n = \frac{1}{d}(\sum_j \ketbra{x_j} + \sum_{j \neq k} \epsilon^n_{jk} \ketbra{x_j}{x_k})$, where $|\epsilon^n_{jk}| \le 1$ for $n = 1,2$. The convex combination $\upsilon = \lambda \upsilon_1 + (1 - \lambda)\upsilon_2$ is also an unpredictable state, since $\expval{\upsilon }{x_j} = 1/d_A \ \forall j.$

Now, to prove that  $\Upsilon_X(\mathcal{H_A})$ is \textit{compact}. We have to remember that any bone-fide predictability measure should be  a \textit{continuous} function \textit{only} of the diagonal elements of the density matrix and vanish only for unpredictable states. For instance, our prototypical measure, defined by Eq. (\ref{eq:pvn}), satisfies the criteria established in Refs. \cite{Durr, Englert}. Therefore, the image set of $\Upsilon_X(\mathcal{H_A})$ under any bona fide predictability measure is closed and has only one element: $\{0\}$. Besides, since any bona fide predictability measure is a continuous function of $\rho_{A diag}$, it is possible to conclude that  $\Upsilon_X(\mathcal{H_A})$ is a closed set \cite{Holmes}. Therefore, for finite dimensional Hilbert spaces, this implies that  $\Upsilon_X(\mathcal{H_A})$ is compact.

Since $\Upsilon_X(\mathcal{H_A})$ is a convex and compact set, by the Hahn-Banach theorem \cite{Holmes}, there exists a Hermitian operator $W$ (known as witness) that separates states $\rho_A \notin \Upsilon_X(\mathcal{H_A})$ from the set of unpredictable states $\Upsilon_X(\mathcal{H_A})$. For instance, using the predictability measure $P^X_{vn}(\rho_A)$ defined by Eq. (\ref{eq:pvn}), a possible witness operator is given by $W = \log_2(I_A/d_A) - \log_2(\rho_{Adiag})$. A trivial verification shows that $\Tr W \upsilon_A = 0 \ \forall \upsilon_A \in \Upsilon_X(\mathcal{H_A})$ and $\Tr W \rho_A = - P^X_{vn}(\rho_A) < 0 \ \forall \rho_A \notin \Upsilon_X(\mathcal{H_A})$.  We observe that this witness can only be constructed for full-rank density matrices.

In order to introduce free operations in the RTP, we have to consider a quantum operation (CPTP map) that suppresses all the resource available in the quantum state $\rho_A$, according to Ref. \cite{Lloyd}. In our case, this quantum operation corresponds to two NRvNMs of observables $X,Y$ that are maximally incompatible, denoted by $\Phi_{XY} = \Phi_{YX}$, as defined in Sec. \ref{sec:pred}. It is easy to see that $P^X_{vn}(\Phi_{XY}(\rho_A)) = 0,$ for any density operator $\rho_A$. Then, it is possible to introduce a quantum operation $\Lambda_{\epsilon}$ referred to as monitoring \cite{Dieguez} and defined by 
\begin{align}
    \Lambda_{\epsilon}(\rho_A) := (1 - \epsilon) \rho_A + \epsilon \Phi_{XY}(\rho_A), \label{eq:lamb}
\end{align}
where the parameter $\epsilon \in [0,1]$ is introduced to characterize the capability of the given map to destroy the resource \cite{Costa}, once $P(\Lambda_0(\rho_A)) = P(\rho_A)$ and $P(\Lambda_1(\rho_A)) = 0$, for any bona fide measure of predictability $P$. We could as well introduce this map as $\Lambda_{\epsilon} = (1 - \epsilon) \Phi_X(\rho_A) + \epsilon \Phi_{XY}(\rho_A)$, once $P$ should be a function only of the diagonal elements of $\rho_A$. In addition, since $P$ should be a convex function of $\rho_A$ \cite{Durr, Englert}, it follows that
\begin{align}
    P(\Lambda_{\epsilon}(\rho_A)) \le P(\rho_A),
\end{align}
where the equality holds only for $\epsilon = 0$. Besides, if $\upsilon_A \in \Upsilon_X(\mathcal{H_A})$, then $  P(\Lambda_{\epsilon}(\upsilon_A)) = 0$. Therefore, as noted in Ref. \cite{Costa}, $\Lambda_{\epsilon}$ is a generic description of a free operation. By the Stinespring theorem, such operation can be implemented through an entangling dynamics between the system and an ancilla, plus future discarding of the ancilla. Other examples of free operations are $e^{-i\xi F}$ with $[F,X]=0$ and also the asymptotic limit of the generalized depolarizing channel \cite{Arsen}. Besides, it is possible to introduce a free operation that always preserves the predictability; we just have to consider a unique non-selective measure of a given observable $X$. By defining the following map
\begin{align}
     \Theta_{\epsilon}(\rho_A) := (1 - \epsilon) \rho_A + \epsilon \Phi_{X}(\rho_A),
\end{align}
it is straightforward to see that $ P(\Theta_{\epsilon}(\rho_A)) = P(\rho_A)$.

It is interesting noticing that
\begin{align}
    \Lambda_{\epsilon} \circ \Phi_{XY} = \Phi_{XY} \circ \Lambda_{\epsilon} \label{eq:comm},
\end{align}
which is the commuting condition, as introduced in Ref. \cite{Lloyd}. This implies that: (i) $\Lambda_{\epsilon} \circ \Phi_{XY} = \Phi_{XY} \circ \Lambda_{\epsilon} \circ \Phi_{XY}$, which shows that our fixed point is $I_A/d_A$ and $\Lambda_{\epsilon}$ is unital; (ii) $\Phi_{XY} \circ \Lambda_{\epsilon}= \Phi_{XY} \circ \Lambda_{\epsilon} \circ \Phi_{XY}$, which means that $\Lambda_{\epsilon}$ cannot make use of the resource stored in any $\rho_A$ to affect the free part. The commuting condition has a special role, since any resource theory with free operations satisfying the commuting  condition has a class of computationally easy monotones, once we can escape the optimizations procedures. We just have to consider a function $D(\rho_A, \sigma_A)$ defined over the two states $\rho_A$ and $\sigma_A$ that is contractive, i.e., $D(\Lambda_{\epsilon}(\rho_A), \Lambda_{\epsilon}(\sigma_A))  \le D(\rho_A, \sigma_A)$, where $D$ is not necessarily a metric. Therefore, $\mathcal{D}(\rho_A) := D(\Phi_X(\rho_A), \Phi_{XY}(\rho_A))$ is a predictability monotone, according to Ref. \cite{Lloyd}, once that $\mathcal{D}(\rho_A) \ge \mathcal{D}(\Lambda_{\epsilon}(\rho_A))$. It is important to observe that in the first argument of $D$, the input is $\Phi_X(\rho_A)$ and not $\rho_A$, once any predictability measure should be a function only of $\rho_{Adiag}$, as we discussed before. For instance, $P^X_{vn}(\rho_A) = S_{vn}(\Phi_X(\rho_A)||\Phi_{YX}(\rho_A))$ satisfies this definition. As well, by remembering that the linear entropy can be obtained from the von Neumann entropy as the first order expansion:  $ - \log \rho \approx 1 - \rho$, it is possible to define the relative linear entropy from the relative entropy as $S_l(\rho||\sigma):= \Tr \Big(\rho(\rho - \sigma)\Big)$, from which we can define the linear predictability $P^X_{l}(\rho_A) := S_l(\Phi_X(\rho_A)||\Phi_{YX}(\rho_A)) = S_l(\rho^X_{Adiag}|| I_A/d_A) = S^{\max}_l - S_l(\rho^X_{Adiag})$, which also satisfies all the criteria established in Refs. \cite{Durr, Englert}, as already shown by us in Ref. \cite{Maziero}. In addition, non symmetric distances such as relative R\'enyi entropies are also valid choices for $D$ \cite{Lloyd}. 

Even though, one can define predictability monotones from the framework above, it is also possible to define them through a minimization procedure over the set of unpredictable states. For instance,
predictability monotones can be defined as
 $$P^X_{vn}(\rho_A) = \min_{\upsilon \in \Upsilon_X(\mathcal{H}_A)} S_{vn}(\rho^X_{Adiag}||\upsilon_A),$$ with the minimum being obtained for $\upsilon_A = I_A/d_A$.
Let $P := \min_{\upsilon_A \in \Upsilon_X(\mathcal{H}_A)}S_{vn}(\rho^X_{Adiag}||\upsilon_A)$. Therefore, we have to show that $P = P^X_{vn}(\rho_A) $. Now,
\begin{align}
    P & = \min_{\upsilon_A \in \Upsilon_X(\mathcal{H}_A)}S_{vn}(\rho^X_{Adiag}||\upsilon_A) \label{eq:min}\\ 
    &  = \min_{\upsilon_A \in \Upsilon_X(\mathcal{H}_A)}\Tr \Big( \rho^X_{Adiag} \log_2 \rho^X_{Adiag} - \rho^X_{Adiag} \log_2 \upsilon_A \Big) \\
    & = \Tr \rho^X_{Adiag} \log_2 \rho^X_{Adiag} -  \max_{\upsilon_A \in \Upsilon_X(\mathcal{H}_A)} \Tr \rho^X_{Adiag} \log_2 \upsilon_A. 
\end{align}
Regarding the maximization of $\Tr \rho^X_{Adiag} \log_2 \upsilon_A, \ \forall \upsilon_A \in \Upsilon_X(\mathcal{H}_A)$, we will show that it is obtained with $\upsilon_A = I_A / d_A$. First, let us notice that
\begin{align}
     \max_{\upsilon \in \Upsilon_X(\mathcal{H}_A)} \Tr \rho^X_{Adiag}& \log_2 \upsilon_A = \\ & \max_{\upsilon \in \Upsilon_X(\mathcal{H}_A)} \sum_k \rho_{kk} \expval{\log_2 \upsilon_A}{k}, \nonumber
\end{align}
where $\ket{k}:=\ket{x_k}$ and $\rho_{kk} = \expval{\rho_A}{x_k}$. Once $\expval{\log_2 \upsilon_A}{k} \le \log_2 (\upsilon_{kk})$, with $\upsilon_{kk} = \expval{\upsilon_A}{x_k}$, the maximization is attained when $\expval{\log_2 \upsilon_A}{k} = \log (\upsilon_{kk})$. Therefore $\upsilon_A$ is diagonal in the reference basis. By imposing that $\Tr \upsilon = 1$ and $\upsilon_{kk} = 1/d_A \ \forall k$, we see that the unpredictable state that minimizes Eq.(\ref{eq:min}) is $I_A/d_A$. With this we shall have
\begin{align}
     P & =  \Tr \rho^X_{Adiag} \log_2 \rho^X_{Adiag} -   \Tr \rho^X_{Adiag} \log_2 I_A/d_A \nonumber \\
     & = S_{vn}(\rho_{diag}||I_A/d_A)\\
     & = P^X_{vn}(\rho_A).
\end{align}


\subsection{Geometrical intuition}
\label{sec:geo}

In this section, for geometrical intuition purposes, we explore the RTP for a qudit using the generalized Gell-Mann's matrices (GMM) and the linear predictability $P^X_{l}(\rho_A)$ as the first order aproximation of $P^X_{vn}(\rho_A)$  . Let us start with a qubit. It is well known that in this case the density operator can be decomposed as
\begin{align}
    \rho_A = \frac{1}{2}(I_A + \vec{r} \cdot \vec{\sigma}),
\end{align}
where $\vec{\sigma} = (\sigma_x, \sigma_y, \sigma_z)$ are the Pauli matrices and $\vec{r} = (r_x, r_y, r_z)$ is a vector in $\mathbb{R}^3$ that uniquely determine the density operator $\rho_A$. The set of density operators for qubits can be represented by the Bloch ball $\mathcal{B}(\mathbb{R}^3) = \{\vec{r} \in \mathbb{R}^3 \ \text{such that} \ \abs{\vec{r}} \le 1\}$ \cite{Kimura}.

\begin{figure}[t]
    \centering
    \includegraphics[scale=0.38]{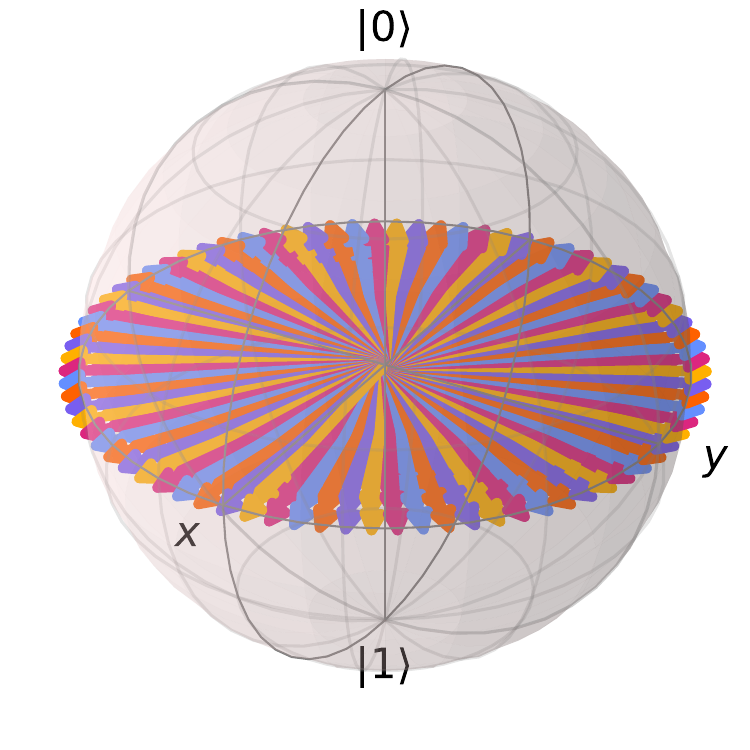}
    \caption{Illustration of the free states (the colorized equatorial plane) and maximally resourceful states (the poles) in the resource theory of predictability of qubit states with reference to the observable $\sigma_{z}$.}
    \label{fig:blockb}
\end{figure}

For $X=\sigma_{z}$, one can easily see that the set of unpredictable states $\Upsilon_X(\mathcal{H}_A)$ is represented by the intersection between the plane defined by $r_z = 0$ and the Bloch ball, i.e., $\Upsilon_X(\mathcal{H}_A)$ is the equatorial close disk of the Bloch ball. By performing a NRvNM of $X$, the state $\rho_A$ is projected to the $r_z$ axis, i.e., $\Phi_X(\rho_A) = \frac{1}{2}(I_A + r_z \sigma_z)$. Thus, one can be seen from  Fig. \ref{fig:blockb} that the closest unpredictable state of $\Phi_X(\rho_A)$ is the maximally mixed state, which is the origin of the Bloch ball. This suggests a natural definition for the predictability measure as
\begin{align}
    P(\rho_A) = \abs{r_z} = \abs{\rho_{11} - \rho_{22}} = \abs{\Tr (\rho_A \sigma_z)},
\end{align}
where $\rho_{jj} = \expval{\rho_A}{x_j}$. In fact, this is a well known predictability measure for qubits, introduced in Ref. \cite{Yasin} and used very often in the literature \cite{Janos}. Besides, it is noteworthy that, in this case, $ P^X_l(\rho_A) = \frac{1}{2} P^2$, where $P_l(\rho_A)$ is the linear predictability. 

Now, let us consider a qudit state $\rho_A$ and an observable $X$. Using the eigenbasis of $X$, $\{\ket{x_j}\}_{j=1}^{d_A},$ we can define the generalized Gell-Mann's Matrices (GMM) as \cite{Krammer}:
\begin{align}
\Gamma_{m}^{d} & :=  \sqrt{\frac{2}{m(m+1)}}\sum_{l=1}^{m+1}(-m)^{\delta_{l,m+1}}\ketbra{x_l},\\
\Gamma_{j,k}^{s} & :=  \ketbra{x_j}{x_k} + \ketbra{x_k}{x_j},\\
\Gamma_{j,k}^{a} & :=  -i(\ketbra{x_j}{x_k} - \ketbra{x_k}{x_j}),
\end{align}
where $ m=1,\cdots,d-1\mbox{ and }1\le j<k\le d$. For $d = 2$, the GMM are the well known Pauli matrices. By defining $I_A :=\Gamma_{0}^{d}$, one can see that under the Hilbert-Schmidt's inner product, $\langle A|B\rangle_{hs}:=\mathrm{Tr}(A^{\dagger}B),$ with $A,B\in\mathbb{C}^{d \times d}$, the set $\left\{\frac{\Gamma_{0}^{d}}{\sqrt{d}},\frac{\Gamma_{m}^{d}}{\sqrt{2}},\frac{\Gamma_{j,k}^{\tau}}{\sqrt{2}}\right\},$ with $\tau=s,a$, forms an orthonormal basis for $\mathbb{C}^{d \times d}$. Thus, we can decompose  the density operator as follows
\begin{equation}
\rho_A =\frac{1}{d}\Gamma_{0}^{d}+\frac{1}{2}\sum_{m}\langle\Gamma_{m}^{d}|\rho\rangle\Gamma_{m}^{d}+\frac{1}{2}\sum_{k,l,\tau}\langle\Gamma_{k,l}^{\tau}|\rho\rangle\Gamma_{k,l}^{\tau},
\end{equation}
while the unpredictable states are of the form
\begin{align}
    \upsilon_A =\frac{1}{d}\Gamma_{0}^{d}+\frac{1}{2}\sum_{k,l,\tau}\langle\Gamma_{k,l}^{\tau}|\rho\rangle\Gamma_{k,l}^{\tau},
\end{align}
where $\langle\Gamma_{m}^{d}|\rho\rangle = 0 \ \forall m$. By performing a NRvNM of $X$, the state $\rho_A$ is projected into the incoherent state
\begin{equation}
    \Phi_X(\rho_A) = \frac{1}{d}\Gamma_{0}^{d}+\frac{1}{2}\sum_{m}\langle\Gamma_{m}^{d}|\rho\rangle\Gamma_{j}^{d},
\end{equation}
such that the linear predictability can be expressed as
\begin{equation}
    P^X_l(\rho_A) = \frac{1}{2}\sum_m \langle\Gamma_{m}^{d}|\rho\rangle^2 = \frac{1}{d} \sum_{j < k} (\rho_{jj} - \rho_{kk})^2.
\end{equation}

\subsection{Composite systems}
\label{sec:composite}
In this section, we discuss our prototypical predictability measure for the bipartite case. If we consider a bipartite quantum system $AB$ describe by $\rho_{AB}$, it is possible to write the joint predictability as
\begin{align}
    P_{vn}(\rho_{AB}) & = S_{vn}(\rho_{AB_\text{diag}}||I_{AB}/d_{AB}) \nonumber \\ & = \log_2 d_A d_B - S_{vn}(\rho_{AB_\text{diag}}) \nonumber \\
    &= P_{vn}(\rho_{A}) + P_{vn}(\rho_{B}) + I_{A:B}(\Phi_{XY}(\rho_{AB})), 
\end{align}
where $I_{A:B}(\Phi_{XY}(\rho_{AB}))$ is the mutual information of $\Phi_{XY}(\rho_{AB}) = \rho_{AB_\text{diag}}$, and is measuring the classical correlation between $A$ and $B$ given that we perform a  non-revealing measurement of the observables $X$ and $Y$. Now, $X$ is an observable of the subsystem $A$, while $Y$ is an observable of the subsystem $B$ and the predictability measure $P_{vn}(\rho_{AB}) $ is written in terms of the eigenbasis of the observables $X$ and $Y$. Therefore, one can see that, if the bipartite system is uncorrelated, i.e., if $\rho_{AB} = \rho_A \otimes \rho_B$, then $P_{vn}(\rho_{AB}) = P_{vn}(\rho_{A}) + P_{vn}(\rho_{B})$, which is to be expected. Besides, $P_{vn}(\rho_A^{\otimes n}) = n P_{vn}(\rho_A)$ once that $S_{vn}(\rho^{\otimes n}|| \sigma^{\otimes n}) = nS_{vn}(\rho||\sigma)$. On the other hand, if we have a maximally entangled state of two qudits, then $P_{vn}(\rho_A) = P_{vn}(\rho_B) = 0$ and $P_{vn}(\rho_{AB}) = I_{A:B}(\Phi_{XY}(\rho_{AB})) = \log_2 d$. Therefore, when we have a maximally entangled state of two qudits and perform non-selective measures in both parts, the maximum information that we can extract from the system is $\log_2 d$ bits of information. Finally, it is noteworthy that the joint predictability $P_{vn}(\rho_{AB})$ is equal to the bipartite quantum incompatibility measure,  introduced in Ref. \cite{Savi}, for MU observables in both parts of system.

Besides, given the joint predictability, it is possible to define that conditional predictability of $A$ given $B$ as
\begin{align}
    P_{vn}(\rho_{A|B}) & := P_{vn}(\rho_{AB}) - P_{vn}(\rho_{B}) \\
    & = S_{vn}(\rho_{AB_\text{diag}} || \frac{I_A}{d_A} \otimes \rho_{B_\text{diag}}), 
\end{align}
which can be rewritten as
\begin{align}
    P_{vn}(\rho_{A|B}) = P_{vn}(\rho_{A})  + I_{A:B}(\Phi_{XY}(\rho_{AB})).
\end{align}
Therefore, the conditional predictability of $A$ given $B$ can be interpreted as the predictability that we already have about $A$ plus the classical correlation between $A$ and $B$. The above equation can still be rewritten as
\begin{align}
     P_{vn}(\rho_{A|B}) & = \log_2 d_A -  S_{A|B}(\Phi_{XY}(\rho_{AB})) \\
     &= I_{A|B}(\Phi_{XY}(\rho_{AB})),
\end{align}
where $S_{A|B}(\Phi_{XY}(\rho_{AB}))$ is the quantum conditional entropy of $\Phi_{XY}(\rho_{AB}) = \rho_{AB_\text{diag}}$, which in this case is always positive, because $\rho_{AB_\text{diag}}$ is a classical state, and $I_{A|B}(\Phi_{XY}(\rho_AB))$ is the conditional information about the state $\rho_{AB_\text{diag}}$. Besides, we have the following expected complementarity relation $   P_{vn}(\rho_{A|B}) +  S_{A|B}(\Phi_{XY}(\rho_{AB})) = \log_2 d_A. $

\subsection{Relation among the resource theories of predictability, coherence, and purity}
\label{sec:RTs}
In this section, we will discuss the relation among the resource theories of predictability, coherence and purity within the resource destroying maps-based framework developed in Ref. \cite{Lloyd} and on the unified approach taken in Ref. \cite{Costa}. The set of incoherent states (free states in the resource theory of quantum coherence) with reference to the observable $X$, $\mathcal{I}_X(\mathcal{H_A})$, is given by states of the type $\Phi_X(\rho_A) = \sum_j X_j \rho_A X_j$ such that the free operations can be recast in terms of the coherence destroying monitoring operation $\Theta_{\epsilon}(\rho_A) = (1 - \epsilon) \rho + \epsilon \Phi_{X}(\rho),$ already introduced before. By noticing that $\Theta_{\epsilon} \circ \Phi_X = \Phi_X \circ \Theta_{\epsilon}$, the coherence monotone is given by $C^X_{re}(\rho_A) = S_{vn}(\rho_A||\Phi_X(\rho_A))$, which is just the relative entropy of quantum coherence. Meanwhile, the set maximally incoherent states (the free states of the resource theory of purity) 
with reference to the observable $X$, $\mathcal{M}_X(\mathcal{H_A})$, is composed of only one element: $I_A/d_A$, which is obtained by the maximally destroying map $\Phi_{XY}$ where $X,Y$ are MU, as we introduced before. The free operations are just $\Lambda_{\epsilon} =  (1 - \epsilon) \rho + \epsilon \Phi_{XY}(\rho)$ such that, from the commuting condition given by Eq.(\ref{eq:comm}), we have the well known (local) information measure (or purity measure) $I(\rho_A) = S_{vn}(\rho_A|| \Phi_{XY}(\rho_A)) = \log_2 d_A - S_{vn}(\rho_A)$ \cite{Hor}.

Therefore, one can easily see that $\mathcal{M}_X(\mathcal{H_A}) = \mathcal{I}_X(\mathcal{H_A}) \cap \Upsilon_X(\mathcal{H_A}).$ Besides, the sum of the complementarity measures gives us the information measure
\begin{align}
  C^X_{re}(\rho_A) + P^X_{vn}(\rho_A) = I(\rho_A) \le \log_2 d_A.
\end{align}
Now, from an operational point of view, by applying non-selective measures of MU observables $X,Y$, we can see which type of resource is erased from the state $\rho_A$. First, by performing NRvNM of $X$, expressed by the map $\Phi_X(\rho_A)$, the quantum coherence with respect to $X$ is destroyed, while the predictability is left intact. This implies that the information content of the state $\rho_A$ has decreased: $I(\Phi_X(\rho_A)) \le I(\rho_A)$, once $C^X_{re}(\Phi_X(\rho_A)) = 0$. Then, by performing a NRvNM of $Y$, we destroy the predictability of the state and completely erase the informational content of $\rho_A$, once $P^X_{vn}(\Phi_{YX}(\rho_A)) = I(\Phi_{YX}(\rho_A)) = 0$.

Finally, it is interesting to see that the complementarity relation given by Eq. (\ref{eq:cr1}) saturates if and only if $\rho_A$ is a pure state, as is the case with any complementarity relation derived in Ref. \cite{Maziero}. As well, predictability and coherence measures always vanish for maximally mixed states. By considering the free operation given by Eq. (\ref{eq:lamb}), one can see that a resource theory of complementarity is equivalent to the resource theory of purity. Besides, complementarity relations, that saturates if and only if the state of the quanton is pure and vanish for maximally mixed states, can be used as purity measures, as introduced in Ref. \cite{Mauro}.

\section{Conclusions}
\label{sec:con}
In this work, we addressed the question if the predictability can be taken as a quantum resource. First, we showed that the predictability with reference to an observable $X$ is equal to the quantum coherence of the state $\Phi_X(\rho_A)$ with reference to observables mutually unbiased (MU) to $X$. As well, we extended this result to observables not necessarily MU and showed that $P^X_{vn}(\rho_A)$ dictates how much quantum coherence, with respect to the observable $Y$, we will have encoded in the state $\Phi_X(\rho_A)$, given that we perform a non-revealing von Neumann measurement (NRvNM) of $Y$. Besides, we showed that the predictability is closely related to the quantum incompatibility measure, introduced in Ref. \cite{Martins}, which is another type of quantum resource. Moreover, we provided quantum circuits to implement NRvNM and use these circuits to experimentally test these (in)equalities using the IBM quantum computers. Once quantum discord is the minimal amount of correlations destroyed by local von Neumann non-revealing measurements, we envisage that the quantum circuits introduced here can be useful for experiments in this context as well. What is more, we formally presented a resource theory of predictability by identifying its free states and operations and giving some predictability monotones, as well as discussing its relationship with other well known quantum resource theories.

\begin{acknowledgments}
This work was supported by Universidade Federal do ABC (UFABC), process 23006.000123/2018-23, and by the Instituto Nacional de Ci\^encia e Tecnologia de Informa\c{c}\~ao Qu\^antica (INCT-IQ), process 465469/2014-0. We thank Renato Moreira Angelo for the question we answered in this article and for illuminating conversations. We also thank Alexandre Camacho Orthey Junior for discussions regarding the implementation of local non-revealing von Neumann measurements through global unitaries. 
\end{acknowledgments}


\end{document}